\documentclass[a4paper,12pt]{article}
\usepackage[english]{babel}
\usepackage{amsmath,amsthm}%
\usepackage{cite}%
\newtheorem{theorem}{Theorem}
\newtheorem*{hyp}{Hypothesis}
\newtheorem{coral}{Corollary}
\usepackage{hyperref}
\usepackage{lamsarrow}
\usepackage{pb-diagram,pb-lams} 

\def\eq#1{\begin{equation}#1\end{equation}}
\def\eqs#1{\begin{eqnarray}#1\end{eqnarray}}
\def\eqn#1{\begin{equation}\begin{split}#1\end{split}\end{equation}}
\def\seq#1{\begin{equation*}#1\end{equation*}}
\def\seqs#1{\begin{equation*}\begin{split}#1\end{split}\end{equation*}}
\def\qed{\vrule height0.6em width0.3em depth0pt\medskip}
\def\matrixx#1{\left(\begin{array}{cc}#1\end{array}\right)}

\def\ad{\hbox{ad}}
\def\ourBn{(\ref{our_bN},\ref{bN})}

\def\Z {\hbox{\Sets Z}}
\def\N {\hbox{\Sets N}}
\def\dfrac#1#2{\frac{\partial #1}{\partial #2}}
\newtheorem{lemma}{Lemma}

\newcounter{transf}
\newcounter{meq}

\font\Sets=msbm10

\title{\bf An unusual series of autonomous discrete integrable equations on the square lattice}
\author{{\bf R.N. Garifullin and R.I. Yamilov}
\\ Institute of Mathematics, Ufa Federal Research Centre,\\ Russian Academy of Sciences,\\ 112 Chernyshevsky Street, Ufa 450008, Russian Federation\\
{\sl E-mails: rustem@matem.anrb.ru, RvlYamilov@matem.anrb.ru}}

\begin{document}
\maketitle
\abstract{We present an infinite series of autonomous discrete equations on the square lattice possessing hierarchies of autonomous generalized symmetries and conservation laws in both directions. Their orders in both directions are equal to $\kappa N$, where $\kappa$ is an arbitrary natural number and $N$ is equation number in the series. Such a structure of hierarchies is new for discrete equations in the case $N>2$. 

Symmetries and conservation laws are constructed by means of the master symmetries. Those master symmetries are found in a direct way together with generalized symmetries. Such construction scheme seems to be new in the case of conservation laws. One more new point is that, in one of directions, we introduce the master symmetry time into coefficients of discrete equations.

In most interesting case $N=2$ we show that a second order generalized symmetry is closely related to a relativistic Toda type integrable equation. As far as we know, this property is very rare in the case of autonomous discrete equations.}

\section{ Introduction}
The following discrete integrable equation 
 \eq{(u_{n,m+1}+1)(u_{n,m}-1)=(u_{n+1,m+1}-1)(u_{n+1,m}+1)\label{d_dr_old}} is well-known \cite{ht95,nc95}.
Recently a number of integrable generalizations of this equation have been found \cite{ly09,gy15,ghy15}. All of them are non-autonomous, and here we write down the two most interesting. One of them reads: 
 \eqn{(u_{n,m+1}+\chi_{n+m+1})(u_{n,m}-\chi_{n+m})=(u_{n+1,m+1}-\chi_{n+m})(u_{n+1,m}+\chi_{n+m+1}),\\ \chi_k=\frac12(1+(-1)^k)\label{dw1},} 
and this is the equation \cite[(77)]{gy15} up to the involution $n\leftrightarrow m.$ 

The second example actually represents a series of discrete equations corresponding to some periods of $n$-dependent coefficient. For any fixed $N\geq 1$, an equation is defined by
\eqn{\alpha_n(u_{n,m+1}+1)(u_{n,m}-1)=\alpha_{n+1}(u_{n+1,m+1}-1)(u_{n+1,m}+1),\\ \alpha_{n+N}=\alpha_n\neq0,\quad\hbox{for all } n\in\Z,\label{our}} 
and it has been studied in \cite{ghy15}.
In both cases these generalizations have hierarchies of generalized symmetries and conservations laws in both directions as well as the $L-A$ pairs, but all these objects are non-autonomous, i.e. explicitly depend on the discrete variables $n$ or $m$. 

Here we are going to construct a series of autonomous integrable generalizations of \eqref{d_dr_old}. We show that all equations of that series have autonomous $L-A$ pairs, generalized symmetries and conservations laws.
In particular, that series provides us with examples of autonomous discrete equations, such that the minimal possible orders of their autonomous generalized symmetries in any direction can be arbitrarily high.

A series of equations we construct here is a particular case of \eqref{our}, however, results of the present paper are not a direct consequence of the results presented in \cite{ghy15}. 
 
In Section \ref{sec_auto} we consider an autonomous generalization of \eqref{d_dr_old} with an arbitrary constant coefficient, which includes the whole series under consideration, and construct for it hierarchies of generalized symmetries and of conservation laws in the $m$-direction. Those results are necessary for the next sections.

In Section \ref{sec_int} we construct and study a series of autonomous integrable generalizations of \eqref{d_dr_old} which is the aim of the present work. Autonomous generalized symmetries and conservation laws in the $m$-direction are constructed in Section \ref{sym_m_auto}, while symmetries and conservation laws in the $n$-direction are discussed in Sections \ref{sym_n}, \ref{con_law_n}.
The most interesting case $N=2$ is considered in more detail in Section \ref{sec_N2}, and its relationship with a relativistic Toda type equation is discussed. Autonomous $L-A$ pairs for equations of the series are constructed in Section \ref{sec_la}.
 
Based on the results of the present paper, we formulate and discuss in Conclusion an important hypothesis about the symmetry structure of equations of the series. We also briefly discuss there all the new results obtained.

\section{Autonomous generalization of \eqref{d_dr_old} with an arbitrary constant coefficient} \label{sec_auto}

The most broad generalization of equation \eqref{d_dr_old} we know is 
 \eqn{(u_{n,m+1}+a_{n,m+1})(u_{n,m}-a_{n,m})=(u_{n+1,m+1}-b_{n+1,m+1})(u_{n+1,m}+b_{n+1,m}),\\ a_{n,m+2}=a_{n,m},\quad b_{n,m+2}=b_{n,m},\quad a_{n,m}^2=b_{n,m}^2. \label{bacmV2}} This is equation \cite[(40)]{gy15} up to transformations $n\leftrightarrow m$ and $b_{n,m}\to-b_{n,m}.$
Equations \eqref{d_dr_old} and \eqref{dw1} are its particular cases. In the case $b_{n,m}=a_{n,m}\neq0$ for all $n,m$, after rescaling $u_{n,m}=\hat u_{n,m}a_{n,m}$, we get for the function $\hat u_{n,m}$ the following equation:
\eqn{\alpha_n(u_{n,m+1}+1)(u_{n,m}-1)=\alpha_{n+1}(u_{n+1,m+1}-1)(u_{n+1,m}+1),\quad \alpha_n\neq0,\label{our_n}} with $\alpha_n=a_{n,m+1}a_{n,m}.$ This equation was introduced in \cite[Section 3]{ly09} in a little bit different form.

There is in \eqref{our_n} an obvious autonomous subcase with an arbitrary constant coefficient~$\beta$:
\eqn{(u_{n,m+1}+1)(u_{n,m}-1)=\beta(u_{n+1,m+1}-1)(u_{n+1,m}+1),\quad \beta\neq0.\label{our_b}}
 It corresponds to the restriction $\alpha_{n+1}/\alpha_n=\beta$ for all $n$, i.e. up to a multiplier we get: \eq{\alpha_n=\beta^n.\label{alpha_n}}
This equation possesses an $L-A$ pair and hierarchies of generalized symmetries and conservation laws in the $m$-direction, but all these objects are non-autonomous. Equation \eqref{our_b} includes the whole series of equations which is the aim of this paper. The results we present here are necessary to the next sections.

An $L-A$ pair for equation \eqref{our_b} is given by:
\eq{\Psi_{n+1,m}=L^{(1)}_{n,m}\Psi_{n,m},\quad \Psi_{n,m+1}=L^{(2)}_{n,m}\Psi_{n,m},\label{lax_d_dress}}where
\eqs{&L^{(1)}_{n,m}=\matrixx{1&2\lambda\beta^n(u_{n,m}+1)\\ -\frac2{u_{n,m}-1}&\frac{u_{n,m}+1}{u_{n,m}-1}},\label{L1}\\ \label{L2}&L^{(2)}_{n,m}=\matrixx{1&-\lambda\beta^n(u_{n,m}+1)(u_{n,m+1}-1)\\ 1&0},}  $\Psi_{n,m}$ is the vector-function and $\lambda$ is the spectral parameter. In more general form, corresponding to \eqref{our_n}, it was presented in \cite{ghy15}, while it was constructed for the first time in \cite{ly09}.

\subsection{Generalized symmetries in the $m$-direction }\label{sec_sym_m}
Here we construct the generalized symmetries in the $m$-direction. A differential-difference equation of the form 
\eq{\partial_t u_{n,m}=h_{n,m}(u_{n,m+\mu},u_{n,m+\mu-1},\ldots, u_{n,m-\mu}),\quad \mu>0,\label{s_gen}} is called the generalized symmetry in the $m$-direction of the discrete equation
\eq{\Phi_{n,m}(u_{n,m},u_{n+1,m},u_{n,m+1},u_{n+1,m+1})=0\label{dis_gen}}
if equations \eqref{s_gen} and \eqref{dis_gen} are compatible. The compatibility condition is obtained by differentiating  \eqref{dis_gen} with respect to the time $t$ in virtue of \eqref{s_gen}:
\seq{\sum_{i,j\in \{0,1\}}h_{n+i,m+j}\dfrac{\Phi_{n,m}}{u_{n+i,m+j}}=0\label{def_sym_m},} and it must be identically satisfied on the solutions of \eqref{dis_gen}. 

We suppose that there exist numbers $n_1,$ $m_1,$ $n_2,$ $m_2$, such that
\eq{\dfrac{h_{n_1,m_1}}{u_{n_1,m_1+\mu}}\neq0,\qquad \dfrac{h_{n_2,m_2}}{u_{n_2,m_2-\mu}}\neq0.}The number $\mu$ is called the order of the generalized symmetry \eqref{s_gen}. The form of equation \eqref{s_gen} is symmetric in a sense. An explanation why such a form is natural for integrable differential-difference equations see in \cite[Section~2.4.1]{y06}.

First we discuss the particular case of \eqref{our_b} with $\beta=1$ which is known. It is important, as generalized symmetries for the general case \eqref{our} are constructed in terms of symmetries of this particular case. 

Its simplest generalized symmetry in the $m$-direction is
\eq{\partial_{t_1'} u_{n,m}=(u_{n,m}^2-1)(u_{n,m+1}-u_{n,m-1})=f_{n,m}^{(1)},\label{mV}} and this is nothing but the modified Volterra equation. 
The known master symmetry of \eqref{mV}, see \cite{ztu91}, can be written in the form: \eq{\partial_{\tau'} u_{n,m}=(u_{n,m}^2-1)((m+1)u_{n,m+1}-(m-1)u_{n,m-1})=g_{n,m}.\label{mtau}} The hierarchy of equation \eqref{mV} can be constructed as follows: 
\eq{\partial_{t_k'} u_{n,m}=f_{n,m}^{(k)}(u_{n,m+k},u_{n,m+k-1},\ldots,u_{n,m-k}),\ k\ge 1,\label{hmV}}
\eqn{f_{n,m}^{(k+1)}&=\ad_{g_{n,m}}f_{n,m}^{(k)}=[g_{n,m},f_{n,m}^{(k)}]=D_{\tau'}f_{n,m}^{(k)}-D_{t_k'}g_{n,m}\\&=\sum_{j=-k}^k g_{n,m+j}\dfrac{f_{n,m}^{(k)}}{u_{n,m+j}}-\sum_{j=-1}^1 f_{n,m+j}^{(k)}\dfrac{g_{n,m}}{u_{n,m+j}}.\label{ad_g}}
Here $D_{\tau'}$ and $D_{t_k'}$ are the operators of total derivatives in virtue of equations \eqref{mtau} and \eqref{hmV}, respectively, with a definition shown in \eqref{ad_g}. 

We get in this way the standard and  known symmetries of the modified Volterra equation.
As $[f_{n,m}^{(1)},f_{n,m}^{(k)}]=0$ for so-constructed functions and
\seq{g_{n,m}=mf_{n,m}^{(1)}+(u_{n,m}^2-1)(u_{n,m+1}+u_{n,m-1}),}
it is easy to prove by induction that  all the functions $f_{n,m}^{(k)}$ do not depend on $m$ explicitly, e.g.:
\eq{\label{f2}f_{n,m}^{(2)}=(u_{n,m}^2-1)[(u_{n,m+1}^2-1)(u_{n,m+2}+u_{n,m})-(u_{n,m-1}^2-1)(u_{n,m}+u_{n,m-2})].} It can be shown, see an explanation below, that \eqref{hmV} are the generalized symmetries of the discrete equation \eqref{our_b} with $\beta=1$ too.
We also notice that both \eqref{mV} and its master symmetry \eqref{mtau} are generalized symmetries of the discrete equation \eqref{our_b} with $\beta=1$.

In general case \eqref{our_b} the simplest generalized symmetry in the $m$-direction reads:
\eq{\partial_{t_1} u_{n,m}=\beta^n f_{n,m}^{(1)}.\label{mV1}} Its master symmetry can be taken in the form:
\eq{\partial_{\tau''} u_{n,m}=\beta^n g_{n,m}.\label{mtau1}} But \eqref{mtau1} is not a generalized symmetry of \eqref{our_b} and, therefore, it allows one to construct generalized symmetries for \eqref{mV1}, but not for \eqref{our_b}. To solve this problem, we need to introduce a special dependence on the master symmetry time into the discrete equation \eqref{our_b} and into both equations \eqref{mV1} and \eqref{mtau1}. 

Such a scheme with introducing the time of master symmetry into a discrete equation is used, probably, for the first time.  

Let us consider a special generalization of \eqref{our_b}:
\eq{A_n(\tau)(u_{n,m+1}+1)(u_{n,m}-1)=A_{n+1}(\tau)(u_{n+1,m+1}-1)(u_{n+1,m}+1),\label{our_A}}
where \eq{A_n(\tau)=(\beta^{-n}+4\tau)^{-1},\quad A_n'(\tau)=-4A_n^2(\tau),\quad A_n(0)=\beta^n,\label{An}} and $\tau$ is an external parameter. Here $\tau$ is the time of a master symmetry.  
It can be checked that the both following equations are generalized symmetries of \eqref{our_A}:
\eq{\partial_{t_1} u_{n,m}=F_{n,m}^{(1)}=A_n(\tau)f_{n,m}^{(1)},\label{sym_A}}
\eq{\partial_{\tau} u_{n,m}=G_{n,m}=A_n(\tau)g_{n,m}.\label{ms_A}}
In particular, the important relation $A_n'=-4A_n^2$ of \eqref{An} is the consequence of compatibility of \eqref{our_A} and \eqref{ms_A}.
As \eqref{ms_A} does not commutate with \eqref{sym_A}, it is reasonable to expect that, for any $k\geq1$, the following functions define nontrivial generalized symmetries of \eqref{our_A}:
\eqn{F_{n,m}^{(k+1)}&=\ad_{G_{n,m}}F_{n,m}^{(k)}=[G_{n,m},F_{n,m}^{(k)}]=D_{\tau}F_{n,m}^{(k)}-D_{t_k}G_{n,m}\\&=\dfrac{F_{n,m}^{(k)}}{\tau}+\sum_{j=-k}^k G_{n,m+j}\dfrac{F_{n,m}^{(k)}}{u_{n,m+j}}-\sum_{j=-1}^1 F_{n,m+j}^{(k)}\dfrac{G_{n,m}}{u_{n,m+j}}.\label{ad_G}}

We see that \eqref{ms_A} allows one to construct a hierarchy of generalized symmetries of \eqref{our_A}. It also generates  a hierarchy of conservation laws, see the next section. For this reason equation \eqref{ms_A} plays the role of the master symmetry not only for \eqref{sym_A} but also for the discrete equation \eqref{our_A}.

Now we are going to study the structure of these generalized symmetries in order to extract later autonomous among them.

We can prove by induction that the following formula takes place:
\eq{F_{n,m}^{(k)}=A_n^k(\tau)\sum_{j=0}^{k-1}4^jc_{k,j}f_{n,m}^{(k-j)},\label{str_F}}
where $c_{k,j}$ are some constants, e.g. $$c_{1,0}=1,\quad c_{2,0}=1,\quad c_{2,1}=-1, \quad c_{3,0}=1,\quad c_{3,1}=-3,\quad c_{3,2}=2.$$
We substitute \eqref{str_F} into \eqref{ad_G} and obtain:
\eq{F_{n,m}^{(k+1)}=\dfrac{A_n^k(\tau)}{\tau}\sum_{j=0}^{k-1}4^jc_{k,j}f_{n,m}^{(k-j)}+A_{n}^{k+1}(\tau)\sum_{j=0}^{k-1}4^jc_{k,j}\ad_{g_{n,m}}f_{n,m}^{(k-j)}.}
Taking into account \eqref{ad_g} and \eqref{An}, we get 
\eq{F_{n,m}^{(k+1)}=-kA_n^{k+1}(\tau)\sum_{j=1}^{k}4^{j}c_{k,j-1}f_{n,m}^{(k+1-j)}+A_{n}^{k+1}(\tau)\sum_{j=0}^{k-1}4^jc_{k,j}f_{n,m}^{(k+1-j)}.\label{str_Fk}}
Comparing \eqref{str_F} and \eqref{str_Fk} we derive the following recurrence formulae:
\eq{c_{k+1,j}=c_{k,j}-kc_{k,j-1},\quad c_{k,-1}=c_{k,k}=0,\quad c_{1,0}=1,\quad 0\le j\le k,\quad k\ge 1.\label{rek_c}}

We see that generalized symmetries of \eqref{our_A} have the form:
\eq{\partial_{t_k}u_{n,m}=F_{n,m}^{(k)},\quad  k\ge 1,\label{hier_s}} where the functions $F_{n,m}^{(k)}$ are of the form \eqref{str_F}, while $f_{n,m}^{(k)},A_n(\tau)$ and $c_{k,j}$ are given by (\ref{ad_g},\ref{An}) and \eqref{rek_c}. The order of such a symmetry equals $k$. An explicit dependence on $n$ and $\tau$ is defined by the multiplier $A_n^k(\tau)$, and there is here no explicit dependence on $m$. When $\tau=0$, equation \eqref{our_A} turns into \eqref{our_b} and the symmetries \eqref{hier_s} turn into generalized symmetries of \eqref{our_b}.

\begin{theorem}\label{th_sym}
The discrete equation \eqref{our_b} possesses generalized symmetries of the form
\eq{\partial_{t_k}u_{n,m}=\beta^{nk}\sum_{j=0}^{k-1}4^jc_{k,j}f_{n,m}^{(k-j)},\quad k\ge 1,\label{sym_b}}with $f_{n,m}^{(k)}$ and $c_{k,j}$ defined by (\ref{ad_g},\ref{rek_c}). These symmetries do not depend explicitly on $m$, and a dependence on $n$ is given by the multiplier $\beta^{nk}$.
\end{theorem}

In the case $\beta=1$, we can see that not only the special linear combination of $f_{n,m}^{(k)}$ shown in \eqref{sym_b} defines generalized symmetries of \eqref{our_b}, but also any of the functions $f_{n,m}^{(k)}.$  

\subsection{Conservation laws in the $m$-direction }\label{sec_con_m}
Let us consider the relation 
\eq{(T_n-1)p_{n,m}=(T_m-1)q_{n,m}\label{cl},} where the functions $p_{n,m},q_{n,m}$ depend on $n,m,u_{n+i,m+j}$ and $T_n,T_m$ are the shift operators in the $n$- and $m$-directions: $T_nh_{n,m}=h_{n+1,m},\ T_mh_{n,m}=h_{n,m+1}$. This relation is called the conservation law of the discrete equation \eqref{dis_gen} if \eqref{cl} is identically satisfied on the solutions of \eqref{dis_gen}. By using \eqref{dis_gen} we can rewrite $p_{n,m},q_{n,m}$ in terms of $n,m$ and the functions \eq{\label{dym_v}u_{n+i,m},u_{n,m+j}} only, and we will represent them in such a form. In case of the $m$-direction, $p_{n,m}$ has the form $$ p_{n,m}=p_{n,m}(u_{n,m+k_1},u_{n,m+k_1-1},\ldots, u_{n,m+k_2}),\quad k_1\geq k_2.$$ This function $p_{n,m}$ can be called the conserved density by analogy with the discrete-differential case.

For $k_1> k_2$ we shall obtain conserved densities $p_{n,m}$ such that $$\dfrac{^2p_{n,m}}{u_{n,m+k_1}\partial u_{n,m+k_2}}\neq0$$ for all $n,m$, then the number $k_1-k_2$ can be called the order of this conservation law, see e.g. \cite{y06}. If $k_1=k_2$ and $p_{n,m}$ is not constant, then the conservation law is not trivial, and its order is equal to $0$. Conservation laws of different orders are essentially different.

Conservation laws for \eqref{our_b} are constructed in \cite{hy13} by using the $L-A$ pair (\ref{L1}, \ref{L2}). However, there is there only a way of construction and a few conservation laws. It is difficult to trace on that way the structure of conservation laws and extract autonomous among them.
Here we solve the problem by using the master symmetry \eqref{ms_A}. Such a way of construction of conservation laws is apparently new. 

It is known in the discrete-differential case that, differentiating a conservation law in virtue of the master-symmetry, one obtains new conservation laws, see e.g. \cite{y06}. Here we show that the same is true for the discrete conservation laws \eqref{cl}. We demonstrate this in detail by example of the discrete equation \eqref{our_A} and its master symmetry \eqref{ms_A}. Then, as in previous section, we pass to equation \eqref{our_b} by choosing $\tau=0$.

It is easy to check that the following functions
\eq{\label{qp1}p_{n,m}^{(1)}=A_n(\tau)(u_{n,m+1}-1)(u_{n,m}+1),\quad q_{n,m}^{(1)}=-2A_n(\tau)u_{n,m}} define a conservation law of \eqref{our_A} in the $m$-direction. Using it and the master symmetry \eqref{ms_A}, we can construct a hierarchy of conservation laws for equation \eqref{our_A}:
\eq{(T_n-1)p_{n,m}^{(k)}=(T_m-1)q_{n,m}^{(k)},\quad k\ge 1,\label{pq}} all of which will not depend explicitly on $m$. 
We do this by induction, using the property that
\eq{(T_n-1)D_\tau p_{n,m}^{(k)} =(T_m-1)D_\tau q_{n,m}^{(k)}\label{con_tau}} is also a conservation law, where $D_\tau$ is the total derivative in virtue of the master symmetry \eqref{ms_A}. The master symmetry \eqref{ms_A} is one of generalized symmetries of \eqref{our_A} in the $m$-direction. For this reason, the operator $D_\tau$ commutes with $T_m$ automatically, while it commutes with $T_n$ on solutions of the discrete equation \eqref{our_A}.

New conservation law \eqref{con_tau} depends on $m$ explicitly. To eliminate $m$ we will use the fact that one can add to both sides of the conservation law \eqref{cl} a function of the form \seq{(T_n-1)(T_m-1)h_{n,m}} and get a new conservation law defined by:
\eq{\tilde p_{n,m}=p_{n,m}+(T_m-1)h_{n,m},\quad \tilde q_{n,m}=q_{n,m}+(T_n-1)h_{n,m}.\label{h12}}
Besides, we use the fact that $p_{n,m}^{(k)}$ are also conserved densities for the differential-difference equation \eqref{sym_A}:
\eq{D_{t_1}p_{n,m}^{(k)}=(T_m-1) r_{n,m}^{(k)}\label{pr}.} This is so, as \eqref{pr} is true for $k=1$ with
\seq{r_{n,m}^{(1)}=A_n^2(\tau)(u_{n,m+1}-1)(u_{n,m}^2-1)(u_{n,m-1}+1),} and it is known from the differential-difference case that, if $p_{n,m}^{(k)}$ is a conserved densities of \eqref{sym_A}, then the function $D_\tau p_{n,m}^{(k)}$ is also its conserved density.

Let us suppose that the functions $p_{n,m}^{(k)},\ q_{n,m}^{(k)},\ r_{n,m}^{(k)}$ do not explicitly depend on $m$ for some $k\ge1.$
Then the total derivative $D_\tau p_{n,m}^{(k)}$ has a linear dependence on $m$, namely: \eq{D_\tau p_{n,m}^{(k)}=(m-1)D_{t_1} p_{n,m}^{(k)}+\ldots\label{Dtau}.}
As \eqref{pr} takes place, we can use transformation \eqref{h12} with $h_{n,m}=-(m-1)r_{n,m}^{(k)}$ and get as a result:
\eq{p_{n,m}^{(k+1)}=D_{\tau} p_{n,m}^{(k)}-(T_{m}-1)[(m-1)r_{n,m}^{(k)}],\label{pkp1}} 
\eq{q_{n,m}^{(k+1)}=D_{\tau} q_{n,m}^{(k)}-(T_{n}-1)[(m-1)r_{n,m}^{(k)}].\label{qkp1}}
The function $p_{n,m}^{(k+1)}$ is a new conserved density for the discrete equation \eqref{our_A} and for its symmetry \eqref{sym_A} and it does not explicitly depend on $m$.

Let us explain now  how to construct the function $r_{n,m}^{(k+1)}$ and why it has no explicit dependence on $m$.
We also give a more simple construction scheme for the functions $p_{n,m}^{(k)}$, which provides an important information about their structure, as well as the second more rigorous justification of why these functions are conserved densities of \eqref{sym_A}.

The function 
\eq{v_{n,m}=A_n(\tau)(u_{n,m+1}-1)(u_{n,m}+1)\label{vu}}
satisfies the equations
\eqs{\partial_{t_1}v_{n,m}&=&v_{n,m}(v_{n,m+1}-v_{n,m-1}),\label{Vol}\\
\partial_{\tau}v_{n,m}&=&v_{n,m}((m+2)v_{n,m+1}+v_{n,m}-(m-1)v_{n,m-1}).} This is nothing but the Volterra equation and its master-symmetry \cite{ozf89}. Relation \eqref{vu} is a slight non-autonomous generalization of the well-known discrete Miura transformation. It transforms the problem of construction of $p_{n,m}^{(k)}$ and $r_{n,m}^{(k)}$ into the well-known problem for the Volterra equation. 
In particular, the initial conserved density $p_{n,m}^{(1)}$ takes the form $p_{n,m}^{(1)}=v_{n,m}$ and it is the common density for all generalized symmetries of the Volterra equation \eqref{Vol}. For this reason, it can be strictly proved that for all $k$ the functions $D_\tau^k p_{n,m}^{(1)}$ are conserved densities for \eqref{Vol}, see \cite[Theorem 20]{y06}. 

The function $r_{n,m}^{(1)}$ takes the form $r_{n,m}^{(1)}=v_{n,m}v_{n,m-1}$. All the functions $p_{n,m}^{(k)}$ and $r_{n,m}^{(k)}$ can also be expressed in terms of $v_{n,m+j}$, i.e. relations \eqref{pr} turn into conservation laws of the Volterra equation \eqref{Vol}. The structure of these conservation laws is described by the following lemma:

\begin{lemma}\label{lem1}
For any $k\geq1$, the function $p_{n,m}^{(k)}$ is an autonomous and homogeneous polynomial of the degree $k$ and it is of the form:  \eq{\label{pv}p_{n,m}^{(k)}=P^{(k)}(v_{n,m},v_{n,m+1},\ldots,v_{n,m+k-1}), \quad\dfrac{^2P^{(k)}}{v_{n,m}\partial v_{n,m+k-1}}\neq0.}
The function $r_{n,m}^{(k)}$ is also an autonomous and homogeneous polynomial of degree $k+1$ and it is of the form: 
\eq{\label{rv}r_{n,m}^{(k)}=R^{(k)}(v_{n,m-1},v_{n,m},v_{n,m+1},\ldots,v_{n,m+k-1}).}
\end{lemma}
Let us recall that, for any $k\geq1$, such two functions define a conservation law of order $k-1$ for the differential-difference equation \eqref{Vol}, see e.g. \cite{y06}. The functions $p_{n,m}^{(k)}$ and $r_{n,m}^{(k)}$ are autonomous in the sense that they do not explicitly depend neither on $n$ nor on $m$.

\paragraph{Sketch of proof.} Lemma \ref{lem1} is true for $k=1$. Let us suppose that it is true for a number $k\geq 1$ and prove it for $k+1$.
We will use the same formula \eqref{pkp1} for the construction of $p_{n,m}^{(k+1)}$. In this case, one easily can check that this function satisfies the assertions of the lemma. The function $p_{n,m}^{(k+1)}$ is the next conserved density of \eqref{Vol}. So there exists a function $r_{n,m}^{(k+1)}$ satisfying the relation \eqref{pr} and depending on the functions $v_{n,m+j}$. It easily can be constructed directly from \eqref{pr}, see e.g. \cite{y06}. Moreover, the left hand side of \eqref{pr} is an autonomous homogeneous polynomial of $v_{n,m+j}$ of the degree $k+2$.  If we look for $r_{n,m}^{(k+1)}$ as a homogeneous polynomial, then it exists and is unique and autonomous. The resulting function satisfies Lemma \ref{lem1}. \qed

If in both functions $p^{(k)}_{n,m}$ and $r^{(k)}_{n,m}$ we replace the functions $v_{n,m+j}$ by $u_{n,m+j}$, using \eqref{vu}, we get a conservation law for the symmetry \eqref{sym_A}, and its order will be $k$, see \cite[Theorem 18]{y06}. It is clear that the so-constructed functions $p^{(k)}_{n,m}$ and $r^{(k)}_{n,m}$ do not explicitly depend on $m$. As the function $p^{(k)}_{n,m}$ in \eqref{pv} is the homogeneous polynomial of degree $k$, then its structure in terms of $u_{n,m+j}$ is: 
\eq{p_{n,m}^{(k)}=A_n^k(\tau)\hat P^{(k)}(u_{n,m},u_{n,m+1},\ldots,u_{n,m+k}),\label{pu}} 
where $\hat P^{(k)}$ is an autonomous polynomial. The dependence on $n$ and $\tau$ is defined here by the multiplier $A_n^k(\tau)$ only.

Now we can show that the function $q_{n,m}^{(k+1)}$ has no explicit dependence on $m$, and the structure of $q_{n,m}^{(k)}$ is similar to \eqref{pu}.

\begin{lemma}\label{lem2}
For any $k\geq1$, the function $q_{n,m}^{(k)}$ is  of the form  \eq{\label{qu}q_{n,m}^{(k)}=A_n^k(\tau)\hat Q^{(k)}(u_{n,m},u_{n,m+1},\ldots,u_{n,m+k-1}), }
where $\hat Q^{(k)}$ is an autonomous polynomial. The function $q_{n,m}^{(k+1)}$ can be constructed by the following recurrence formula:
\eq{q_{n,m}^{(k+1)}=\dfrac{q_{n,m}^{(k)}}{\tau}+A_n(\tau)\sum_{j=0}^{k-1}(u_{n,m+j}^2-1)((j+2)u_{n,m+j+1}-ju_{n,m+j-1})\dfrac{q_{n,m}^{(k)}}{u_{n,m+j}}.\label{qk1r}}
\end{lemma}

\paragraph{Proof.}
It follows from relations \eqref{Dtau} and \eqref{qkp1} that the function $q_{n,m}^{(k+1)}$ has a linear dependence on $m$:
$$ q_{n,m}^{(k+1)}=(m-1) W_{n,m}^{(k)}+Z_{n,m}^{(k)},\quad W_{n,m}^{(k)}=D_{t_1}q_{n,m}^{(k)}-(T_n-1)r_{n,m}^{(k)}.$$
Relation \eqref{pq} with $k$ replaced by $k+1$ and the fact that $p^{(k+1)}_{n,m}$ does not depend on $m$ imply that $(T_m-1)W_{n,m}^{(k)}=0$ on the solutions of \eqref{our_A}. 

The  function $W_{n,m}^{(k)}$ can be expressed in terms of $n,\tau$ and $u_{n,m+j}$ only. This is obvious for the function $D_{t_1}q_{n,m}^{(k)}$ and it is true for $r_{n,m}^{(k)}$ in virtue of \eqref{vu} and \eqref{rv}. Definition \eqref{vu} of $v_{n,m}$ and the discrete equation \eqref{our_A} imply
\eq{v_{n+1,m}=A_{n}(\tau)(u_{n,m+1}+1)(u_{n,m}-1),\label{vp1}} therefore, the function 
\seq{T_nr_{n,m}^{(k)}=R^{(k)}(v_{n+1,m-1},v_{n+1,m},\ldots,v_{n+1,m+k-1})} also can be expressed so. It is important that the dependence on $u_{n,m+j}$ in $W_{n,m}^{(k)}$ is polynomial.

Such a function $W_{n,m}^{(k)}$ satisfies $(T_m-1)W_{n,m}^{(k)}=0$ if and only if it does not depend on $u_{n,m+j}$, i.e. $W_{n,m}^{(k)}=\eta_n^{(k)}(\tau)$. This function equals zero if $u_{n,m+j}=1$ for all $j$, therefore, $W_{n,m}^{(k)}\equiv 0.$

Now we get for $q_{n,m}^{(k+1)}$ the formula $q_{n,m}^{(k+1)}=(D_\tau-(m-1)D_{t_1})q_{n,m}^{(k)}$ which can be rewritten as \eqref{qk1r}. The structure \eqref{qu} for $q_{n,m}^{(k+1)}$ follows from (\ref{An},\ref{qp1}) and the recurrence formula \eqref{qk1r}.\qed

In this way we get the following explicit formulae:
\eq{p_{n,m}^{(1)}=v_{n,m},\quad q_{n,m}^{(1)}=-2A_n(\tau)u_{n,m},\quad r_{n,m}^{(1)}=v_{n,m}v_{n,m-1},\label{pqr1}}
\eqn{p_{n,m}^{(2)}&=v_{n,m}(2v_{n,m+1}+v_{n,m}),\\ q_{n,m}^{(2)}&=-4A_n^2(\tau)(u_{n,m+1}u_{n,m}^2-u_{n,m+1}-2u_{n,m}),\\
r_{n,m}^{(2)}&=2v_{n,m}v_{n,m-1}(v_{n,m+1}+v_{n,m}),\label{pqr2}}
\eqn{p_{n,m}^{(3)}&=2v_{n,m}(3v_{n,m+2}v_{n,m+1}+3v_{n,m+1}v_{n,m}+3v_{n,m+1}^2+v_{n,m}^2),\\ q_{n,m}^{(3)}&=-4A_n^3(\tau)[3(u_{n,m}^2-1)(u_{n,m+2}u_{n,m+1}^2+u_{n,m+1}^2u_{n,m}\\&-u_{n,m+2}-4u_{n,m+1}-5u_{n,m})+16u_{n,m}^3],\\
r_{n,m}^{(3)}&=6v_{n,m}v_{n,m-1}(v_{n,m+2}v_{n,m+1}+2v_{n,m+1}v_{n,m}+v_{n,m+1}^2+v_{n,m}^2),\label{pqr3}} with $v_{n,m}$ given by \eqref{vu}, illustrating the construction scheme described above.

When $\tau=0$, the discrete equation \eqref{our_A} turns into \eqref{our_b} and its conservation laws turn into conservation laws of \eqref{our_b}. As $A_n(0)=\beta^n$, then we get the following result for the conservation laws of \eqref{our_b}:

\begin{theorem}\label{th_con}
For any $k\geq 1$, the discrete equation \eqref{our_b} possesses a conservation law \eqref{pq} of the order $k$ defined by  functions of the form:
\eqs{&p_{n,m}^{(k)}=\beta^{nk}\hat P^{(k)}(u_{n,m},u_{n,m+1},\ldots,u_{n,m+k}),\label{Pu}\\ &q_{n,m}^{(k)}=\beta^{nk} \hat Q^{(k)}(u_{n,m},u_{n,m+1},\ldots,u_{n,m+k-1}),\label{Qu}} where the polynomials $\hat P^{(k)}$ and $\hat Q^{(k)}$ do not explicitly depend neither on $n$ nor on $m$.
\end{theorem}

\section{A series of autonomous integrable generalizations}\label{sec_int}
In Section \ref{sec_auto} we have considered the autonomous discrete equation \eqref{our_b} possessing an $L-A$ pair and  hierarchies of generalized symmetries and conservation laws in the $m$-direction. All those objects are, however, essentially non-autonomous. Symmetries, conservation laws and $L-A$ pairs of autonomous discrete equations we consider here will be autonomous, and those equations will have hierarchies of generalized symmetries and conservation laws in both directions $n$ and $m$. 

It has been shown in \cite{ghy15}  that the discrete equation \eqref{our}, which has the periodic coefficient $\alpha_n$, should have hierarchies of generalized symmetries and conservation laws in both directions $n$ and $m$. In case of conservation laws, this was shown by using an $L-A$ pair. In case of symmetries, we studied some particular cases. 

As we are interested in the autonomous equations, we are going to consider the intersection of equations \eqref{our} and \eqref{our_b}.
It follows from \eqref{alpha_n} that $\beta^N=1.$ So, we will consider the following equations
\eqn{(u_{n,m+1}+1)(u_{n,m}-1)=\beta_N(u_{n+1,m+1}-1)(u_{n+1,m}+1),\label{our_bN}} where $\beta_N^N=1,\  N\geq1$.

In order to separate equations with different numbers $N$, we consider here the primitive roots of unit. It is clear that $\beta_1=1$, and this case is well-known, see \eqref{d_dr_old}.
If $N>1$, then 
\eq{\label{bN} \beta_N^N=1,\qquad \beta_N^j\neq 1 \ \hbox{ for all }\   1\leq j<N. }  In particular, \eqn{\beta_1=1,\quad \beta_2= -1,\quad \beta_3=-\frac12\pm i\frac{\sqrt3} 2,\quad \beta_4=\pm i,\label{our_b1234}} i.e. in the last two cases one has two primitive roots corresponding to the signs $+$ and $-$. For any $N>4$, at least two primitive roots exist, which are given by $\beta_N=\exp(\pm 2i\pi/N)$. So, we consider below the series of equations \eqref{our_bN},  such that $\beta_N$ are the primitive roots of unit.

Currently we know only one similar series of integrable discrete equations \cite{gy12u}. Those equations are Darboux integrable and of the Burgers type, and  the minimal order of their first integrals may be arbitrarily high. Equations of the series \eqref{our_bN} are integrable by the inverse scattering method.

For equation \ourBn with $N=2$ we have $\beta_2=-1$, i.e. it reads: \eqs{(u_{n,m+1}+1)(u_{n,m}-1)=-(u_{n+1,m+1}-1)(u_{n+1,m}+1).\label{our_b2}} This is the most interesting example in the series, as it has real coefficients. It was found in \cite{ggy18}, where the authors searched discrete equations on the square lattice, using as a generalized symmetry five-point differential-difference equations obtained in the recent symmetry classification \cite{gyl17,gyl18}.

\subsection{Autonomous generalized symmetries and conservation laws in the $ m$-direction}\label{sym_m_auto}
Here we construct autonomous generalized symmetries and conservation laws in the $m$-direction  for equations (\ref{our_bN},\ref{bN}), using the results of Section \ref{sec_auto}.

In Theorem \ref{th_sym} we constructed symmetries \eqref{sym_b}, where an explicit dependence on $n$ was given by the multiplier $\beta^{nk}$.
It follows from this theorem that equations (\ref{our_bN},\ref{bN}) have infinitely many autonomous generalized symmetries in the $m$-direction, which are given by \eqref{sym_b} with $k=N,2N,3N,\dots.$

\begin{coral}\label{cor_sym_m} For any $N\geq 2$, the discrete equation (\ref{our_bN},\ref{bN}) has autonomous generalized symmetries in the $m$-direction given by (\ref{sym_b},\ref{ad_g},\ref{rek_c}) with $k=\kappa N,\ \kappa \in\N.$
\end{coral}

 For equation \eqref{our_b2} the simplest autonomous generalized symmetry in the $m$-direction is given by
\eqn{\partial_{t_2}u_{n,m}=c_{2,0}f_{n,m}^{(2)}+4c_{2,1}f_{n,m}^{(1)}.}
We find from \eqref{rek_c} that $c_{2,0}=1,c_{2,1}=-1$ and, using (\ref{mV},\ref{f2}) for the functions $f_{n,m}^{(1)},f_{n,m}^{(2)}$, we get the following explicit formula:
\eqn{\partial_{t_2}u_{n,m}=(u_{n,m}^2-1)\big[(u_{n,m+1}^2-1)(u_{n,m+2}+u_{n,m})-(u_{n,m-1}^2-1)(u_{n,m}+u_{n,m-2})\\-4(u_{n,m+1}-u_{n,m-1})\big].\label{sym_mN2}} This symmetry was first found in \cite{ggy18}.

In Theorem \ref{th_con} we constructed conservation laws for equation \eqref{our_b} given by (\ref{Pu},\ref{Qu}), where an explicit dependence on $n$ was given by the multiplier $\beta^{nk}$.
It follows from this theorem that equations (\ref{our_bN},\ref{bN}) have infinitely many autonomous conservation laws in the $m$-direction, and they are given by (\ref{Pu},\ref{Qu}) with $k=N,2N,3N,\dots.$

\begin{coral}\label{con_m} For any $N\geq 2$, the discrete equation (\ref{our_bN},\ref{bN}) has infinitely many autonomous conservation laws, and their orders are multiples of the number $N$.
\end{coral}

In the case of equation \eqref{our_b2}, the simplest autonomous conservation law, taken from \eqref{pqr2},  has the order 2 and is given by: 
\eqn{p_{n,m}^{(2)}&=v_{n,m}(2v_{n,m+1}+v_{n,m}),\quad v_{n,m}=(u_{n,m+1}-1)(u_{n,m}+1),\\ q_{n,m}^{(2)}&=-4(u_{n,m+1}u_{n,m}^2-u_{n,m+1}-2u_{n,m}).\label{pqr2_2}}

\subsection{Generalized symmetries in the $n$-direction}\label{sym_n}

Let us consider generalized symmetries in the $n$-direction. The discrete equation
\eqref{dis_gen}
has a generalized symmetry in the $n$-direction: 
\eq{\partial_\theta u_{n,m}= \zeta_{n,m}(u_{n+\nu,m},u_{n+\nu-1,m},\ldots, u_{n-\nu,m}),\quad \nu>0,\label{s_gen_n}}
if \eqref{dis_gen} and \eqref{s_gen_n} are compatible, i.e. the equation 
\eq{D_\theta \Phi_{n,m}=\sum_{i,j\in \{0,1\}}\zeta_{n+i,m+j}\dfrac{\Phi_{n,m}}{u_{n+i,m+j}}=0\label{def_sym_n}} is identically satisfied on the solutions of \eqref{dis_gen}. It is natural to suppose that there exist numbers $n_1,$ $m_1,$ $n_2,$ $m_2$, such that
\eq{\label{sym_rh}\dfrac{\zeta_{n_1,m_1}}{u_{n_1+\nu,m_1}}\neq0,\qquad \dfrac{\zeta_{n_2,m_2}}{u_{n_2-\nu,m_2}}\neq0.}The number $\nu$ is called the order of symmetry \eqref{s_gen_n}. The form of equation \eqref{s_gen_n} is symmetric as in Section \ref{sec_sym_m} for the same reason.

In \cite[Section 4]{ghy15} two theorems for equation \eqref{our_n} and its "nondegenerate" symmetries of orders 1 and 2 have been proved. Here we prove analogues theorems for equation \eqref{our_b} and its symmetries of orders 1,2 and 3 without the use of any non-degeneracy conditions. 

\begin{theorem} \label{th_sym_n}The following two statements take place: \begin{itemize}\item[1.] If equation \eqref{our_b} has a generalized symmetry \eqref{s_gen_n} in the $n$-direction of order $N$, such that $1\leq N\leq 3$, then $\beta^N=1$, i.e. equation \eqref{our_b} has the form \eqref{our_bN}; \item[2.] Equation (\ref{our_bN}) with $1\leq N\leq 3$ and $\beta_N$ being a primitive root of unit has a generalized symmetry of the order $N$ and does not have generalized symmetries of lower orders.\end{itemize}
\end{theorem}

{\bf Sketch of Proof.} For the construction of generalized symmetries for the discrete equations, we use a method developed in \cite{ly09,ly11,ggh11}, see \cite{ggh11} for its most advanced version. The compatibility condition \eqref{def_sym_n} is a functional equation for the unknown function $\zeta_{n,m}$. The method allows one to get consequences in the form of partial differential equations for $\zeta_{n,m}$, using so-called annihilation operators introduced in \cite{h05}.
\item[1.]
If equation \eqref{our_b} has a generalized symmetry \eqref{s_gen_n} of the order $N$, with $1\leq N\leq 3$, then the simplest differential consequences of \eqref{def_sym_n} have the form:
\eq{(\beta^N-1)\dfrac{\zeta_{n,m}}{u_{n+N,m}}=0,\quad (\beta^N-1)\dfrac{\zeta_{n,m}}{u_{n-N,m}}=0,\label{prod}} and these relations must be satisfied for all $n,m$. Conditions (\ref{sym_rh},\ref{prod}) imply $\beta^N=1$.
\item[2.] For equation \eqref{our_bN} with $1\leq N\leq 3$ and $\beta_N$ being a primitive root of unit, we look for symmetries of the form \eqref{s_gen_n} with $\nu=N$ and we use no restriction like \eqref{sym_rh}. We find the following generalized symmetries. 

In case $N=1$ it has the form: 
\eq{\partial_{\theta_1} u_{n,m}=(u_{n,m}^2-1)\left(\frac{a_{n+1}}{u_{n+1,m}+u_{n,m}}-\frac{a_{n}}{u_{n,m}+u_{n-1,m}}\right).\label{sym_nN1}}
Here $a_n=b+cn$, where $b, c$ are arbitrary constants. 

In case $N=2$ it is of the form: 
\eqn{\partial_{\theta_2} u_{n,m}=&(u_{n,m}^2-1)(T_n-1)\left(\frac{a_{n+1}(u_{n+1,m}+u_{n,m})}{U_{n,m}}+\frac{a_{n}(u_{n-1,m}+u_{n-2,m})}{U_{n-1,m}}\right),\label{sym_nN2}\\ U_{n,m}=&(u_{n+1,m}+u_{n,m})(u_{n,m}+u_{n-1,m})-2(u_{n,m}^2-1).}
 The function $a_n$ is given by $a_n=b_n+cn$, where $c$ is a constant and $b_{n+2}\equiv b_n$ is an arbitrary two-periodic function on $n$. It can be represented as $b_n=b^{(1)}+b^{(2)}(-1)^n$ with two arbitrary constants $b^{(1)},\ b^{(2)}$. 

In case $N=3$ the generalized symmetry has the form: 
\eqn{\partial_{\theta_3} u_{n,m}=&(u_{n,m}^2-1)(T_n-1)\left(\frac{a_{n+2}V_{n,m}}{U_{n,m}}+\frac{a_{n}W_{n,m}}{U_{n-2,m}}+(T_n+1)\frac{a_{n+1}Z_{n,m}}{U_{n-1,m}}\right)\label{sym_nN3},\\V_{n,m}=&\beta_3^2(u_{n+1,m}^2-1)+u_{n+1,m}(u_{n+2,m}-u_{n-1,m})-u_{n+2,m}u_{n-1,m}+1,\\W_{n,m}=&\beta_3(u_{n-2,m}^2-1)+u_{n-2,m}(u_{n-1,m}+u_{n-3,m})+u_{n-1,m}u_{n-3,m}+1,\\Z_{n,m}=&(u_{n+1,m}+u_{n,m})(u_{n-1,m}+u_{n-2,m}),\\U_{n,m}=&\beta_3^2(u_{n+1,m}^2-1)(u_{n,m}+u_{n-1,m})+\beta_3(u_{n,m}^2-1)(u_{n+2,m}+u_{n+1,m})\\+(&u_{n+1,m}u_{n,m}+1)(u_{n+2,m}+u_{n-1,m})+(u_{n+1,m}+u_{n,m})(u_{n+2,m}u_{n-1,m}+1).} 
Here $\beta_3$ is any of two primitive roots shown in \eqref{our_b1234}. The function $a_n$ is given by $a_n=b_n+cn$, where $c$ is a constant and $b_{n+3}\equiv b_n$ is an arbitrary three-periodic function. It can be represented as $b_n=b^{(1)}+b^{(2)}\beta_3^n+b^{(3)}\beta_3^{2n}$, where $b^{(1)},\ b^{(2)},\  b^{(3)}$ are arbitrary constants. 

We see that such an equation \eqref{our_bN} has a generalized symmetry of the order $N$ in all three cases. We also can see that generalized symmetries of lower orders do not exist in cases $N=2,3$ because the requirements $\dfrac{\zeta_{n,m}}{u_{n+N,m}}\equiv 0$ or $\dfrac{\zeta_{n,m}}{u_{n-N,m}}\equiv 0$ imply that $\zeta_{n,m}\equiv 0.$ $\qed$

In case $N=2$ we have the only primitive root $\beta_2=-1$, and formulae for $b_n$ in cases $N=2$ and $N=3$ are analogous. 
We have the following important consequence of Theorem \ref{th_sym_n} for autonomous equations \eqref{our_bN}:

\begin{coral}\label{cor_sym_n_auto} Any of equations (\ref{our_bN}) with $1\leq N\leq 3$ and $\beta_N$ being a primitive root of unit has an autonomous generalized symmetry of the order $N$, given by (\ref{sym_nN1}-\ref{sym_nN3}) with $a_n\equiv 1$, and does not have autonomous generalized symmetries of lower orders. 
\end{coral}

  These autonomous symmetries exemplify integrable differential-difference equations with one continuous variable $\theta_N$ and one discrete variable $n$, while the parameter $m$ is not essential.
The symmetry \eqref{sym_nN1} with $a_n\equiv 1$ has first been found in \cite{ly09} and it corresponds to a well-known Volterra type integrable equation \cite{y06,y83}. The symmetry \eqref{sym_nN2}  is a particular case of a non-autonomous symmetry of the discrete equation \eqref{our_n} and it was found in \cite{ghy15}. Nevertheless, equations (\ref{sym_nN2}) and (\ref{sym_nN3}) with $a_n\equiv 1$ provide  new examples of autonomous integrable differential-difference equations of the second and third orders. 

If $N=2$, the subcases $a_n\equiv 1$ and $a_n\equiv(-1)^n$ of \eqref{sym_nN2} are compatible, i.e. we have here two commuting generalized symmetries of the second order. When $N=3$, the subcases $a_n\equiv 1$, $a_n\equiv \beta_3^n$ and $a_n\equiv\beta_3^{2n}$ of \eqref{sym_nN3} are compatible, i.e. we have three commuting generalized symmetries of the third order.

Equation \eqref{sym_nN1} with $a_n\equiv n$ is a known master symmetry for the differential-difference equation \eqref{sym_nN1} with $a_n\equiv 1$, see \cite{cy95}. It is important for us here that in all the three cases $N=1,2,3$ the symmetry corresponding to $a_n\equiv n$ plays the role of the master symmetry for discrete equation  \eqref{our_bN} with $\beta_N$ being a primitive root of unit. Compared with Section \ref{sec_sym_m}, these master symmetries are more convenient to use, as they do not depend explicitly on the time of the master symmetry. 

Let us denote by
\eq{\partial_{\hat \theta_N} u_{n,m}=\Xi_{n,m}^{(N)} \label{master_nN}} the generalized symmetry of discrete equation \eqref{our_bN} with $N=2$ or $N=3$ corresponding to $a_n\equiv n$ in (\ref{sym_nN2}) or \eqref{sym_nN3}, which plays the role of the master symmetry. We show how to construct generalized symmetries of higher orders:
\eq{\partial_{\tilde\theta_{k,N}} u_{n,m}=\Upsilon_{n,m}^{(k,N)},\quad k\in\N \label{sym_nN},} starting from symmetries (\ref{sym_nN2}) or \eqref{sym_nN3} with periodic coefficient $a_n\equiv b_n$, which correspond to \eqref{sym_nN} with $k=1$. The order of such a symmetry \eqref{sym_nN} will be equal to $kN$.
The right hand sides of these symmetries are constructed by using the recurrence formula:
\eqn{\Upsilon_{n,m}^{(k+1,N)}&=\ad_{\Xi_{n,m}^{(N)}}\Upsilon_{n,m}^{(k,N)}=D_{\hat\theta_N}\Upsilon_{n,m}^{(k,N)}-D_{\tilde\theta_{k,N}}\Xi_{n,m}^{(N)}\\&=\sum_{j=-kN}^{kN} \Xi_{n+j,m}^{(N)}\dfrac{\Upsilon_{n,m}^{(k,N)}}{u_{n+j,m}}-\sum_{j=-N}^N \Upsilon_{n+j,m}^{(k,N)}\dfrac{\Xi_{n,m}^{(N)}}{u_{n+j,m}}.\label{ad_nN1}}
Here $D_{\hat\theta_N}$ and $D_{\tilde\theta_{k,N}}$ are the total derivatives in virtue of \eqref{master_nN} and \eqref{sym_nN}.

\subsection{Comparison of the case $N = 2$ with a known example. Relation with relativistic Toda type equations}\label{sec_N2}

Let us consider in more detail the discrete equation \eqref{our_b2}. We know the only autonomous example analogous to \eqref{our_b2} from the viewpoint of generalized symmetry structure. It was found in \cite{gy12} and then studied in \cite{gmy14}.

This example reads:
\eq{ u_{n+1,m+1}(u_{n,m}-u_{n,m+1})-u_{n+1,m}(u_{n,m}+u_{n,m+1})+2=0\label{tzit}.}
Its generalized symmetries of the first and second order in the $m$-direction are
\eqs{\partial_{t_1} u_{n,m}&=&(-1)^n\frac{u_{n,m+1}u_{n,m-1}+u_{n,m}^2}{u_{n,m+1}+u_{n,m-1}},\label{s1_tz}\\
\partial_{t_2} u_{n,m}&=&\frac{(u_{n,m+2}-u_{n,m-2})(u_{n,m+1}^2-u_{n,m}^2)(u_{n,m}^2-u_{n,m-1}^2)}{(u_{n,m}+u_{n,m-2})(u_{n,m+1}+u_{n,m-1})^2(u_{n,m+2}+u_{n,m}).}\label{e12m2}}
The simplest symmetry in the $n$-direction has the second order:
\eqn{\label{s2_tz}\partial_{\tilde\theta_2} u_{n,m}=(u_{n+1,m}u_{n,m}-1)(u_{n,m}u_{n-1,m}-1)(b_{n+1}u_{n+2,m} - b_{n}u_{n-2,m}),}
where $b_{n+2}\equiv b_{n}$ is an arbitrary two-periodic function, i.e. $b_n=b^{(1)}+b^{(2)}(-1)^n$ with arbitrary constant coefficients $b^{(1)},\ b^{(2)}.$
In case of the discrete equation \eqref{our_b2}, a symmetry analogous to \eqref{s1_tz} reads:
\eq{\partial_{t_1} u_{n,m}=(-1)^n(u_{n,m}^2-1)(u_{n,m+1}-u_{n,m-1}).\label{sym_m1}}
Generalized symmetries of equation \eqref{our_b2} similar to \eqref{e12m2} and \eqref{s2_tz} are \eqref{sym_mN2} and \eqref{sym_nN2} with $a_n\equiv b_n$.

These autonomous discrete equations \eqref{our_b2} and \eqref{tzit} have hierarchies of autonomous generalized symmetries in both directions. The orders of those autonomous symmetries are even and, as we see from above examples, the simplest autonomous generalized symmetries in both directions have the order 2.

In \cite{gmy14} we showed that the differential-difference equation \eqref{s2_tz} was equivalent to a system of Tsuchida \cite{t02}, see details below. However, in the class of five-point differential-difference equations, this is an interesting integrable example as itself. Equation \eqref{sym_nN2} with $a_n\equiv b_n$ seems a new integrable example of the five-point differential-difference equation.

In \cite{gmy14} we outlined that equation \eqref{s2_tz} was similar to relativistic Toda type equations, see the review articles \cite[Sections 4.2,4.3]{asy00} and \cite[Secion 3.3]{y06}, according to its generalized symmetry properties. In \cite{ghy15} we demonstrated such a relation with relativistic Toda type equations in a more explicit way for a non-autonomous equation. Here, following \cite{ghy15}, we will demonstrate such an explicit relation for the equations \eqref{s2_tz} and \eqref{sym_nN2} with $a_n\equiv b_n$.

Let us first consider the generalized symmetry \eqref{s2_tz}. For any fixed $m$ we introduce
$$v_k=u_{2k,m},\ w_k=u_{2k-1,m},\ \varsigma=b_{2k},\ \eta=b_{2k-1},$$ and rewrite \eqref{s2_tz} in the form of a system:
\eqn{\partial_{\tilde\theta_2}v_{k}=(\eta v_{k+1}-\varsigma v_{k-1})(w_{k+1}v_k-1)(v_kw_{k}-1),\\
\partial_{\tilde\theta_2}w_{k}=(\varsigma w_{k+1}-\eta w_{k-1})(v_{k}w_k-1)(w_kv_{k-1}-1).}This is nothing but the Tsuchida system \cite[(3.13)]{t02}. 
In any of two cases 
\seq{\varsigma=1,\ \eta=0\quad \hbox{ or }\quad \varsigma=0,\ \eta=1,} we introduce 
\seq{U_k=\log v_k\quad \hbox{ or }\quad U_k=-\log w_k} and get in any of these four cases the following relativistic Toda type equation:
\seq{\ddot U_k=\dot U_k\left(\dot U_{k+1}-\dot U_{k-1}-e^{U_{k+1}-U_{k}}+e^{U_{k}-U_{k-1}}\right),}
where we denote $\dot U_k=\partial_{\tilde\theta_2} U_k$. This is a known equation presented in \cite[Section 3.3.4, (Ld3) with $\mu=0,\nu=1$]{y06}.  

Now we consider the symmetry \eqref{sym_nN2} with $a_n\equiv b_n$. 
For any fixed $m$ we introduce $\tilde u_n$:
$$u_{n,m}=\frac{\tilde u_{n}+\tilde u_{n+1}}{\tilde u_{n}-\tilde u_{n+1}}.$$ This transformation is not invertible, but it is linearizable, i.e. not of the Miura type, in the terminology of the theoretical paper \cite{gyl16}. As a result we obtain the following integrable modification of \eqref{sym_nN2} with $a_n\equiv b_n$:
\eqn{\partial_{\tilde\theta_2} \tilde u_{n}&=\frac{(\tilde u_{n+2}-\tilde u_{n})(\tilde u_{n+1}-\tilde u_{n})(\tilde u_{n}-\tilde u_{n-1})}{2(\tilde u_{n+2}\tilde u_{n}+\tilde u_{n+1}\tilde u_{n-1})-(\tilde u_{n+2}+\tilde u_{n})(\tilde u_{n+1}+\tilde u_{n-1})}b_{n+1}\\ &+\frac{(\tilde u_{n+1}-\tilde u_{n})(\tilde u_{n}-\tilde u_{n-1})(\tilde u_{n}-\tilde u_{n-2})}{2(\tilde u_{n+1}\tilde u_{n-1}+\tilde u_{n}\tilde u_{n-2})-(\tilde u_{n+1}+\tilde u_{n-1})(\tilde u_{n}+\tilde u_{n-2})}b_n.}

Now we pass to the notations
$$v_k=\tilde u_{2k},\ w_k=\tilde u_{2k-1},\ \varsigma=b_{2k},\ \eta=b_{2k-1}$$ 
and are led to the following system:
\eqn{\partial_{\tilde\theta_2}v_{k}&=\frac{(v_{k+1}-v_k)(w_{k+1}-v_k)(v_k-w_k)}{2(v_{k+1}v_k+w_{k+1}w_k)-(v_{k+1}+v_k)(w_{k+1}+w_k)}\eta\\ &+\frac{(w_{k+1}-v_k)(v_k-w_k)(v_k-v_{k-1})}{2(w_{k+1}w_k+v_kv_{k-1})-(w_{k+1}+w_k)(v_k+v_{k-1})}\varsigma,\\
\partial_{\tilde\theta_2}w_{k}&=\frac{(w_{k+1}-w_k)(v_k-w_k)(w_k-v_{k-1})}{2(w_{k+1}w_k+v_kv_{k-1})-(w_{k+1}+w_k)(v_k+v_{k-1})}\varsigma\\ &+\frac{(v_k-w_k)(w_{k}-v_{k-1})(w_{k}-w_{k-1})}{2(v_{k}v_{k-1}+w_{k}w_{k-1})-(v_{k}+v_{k-1})(w_{k}+w_{k-1})}\eta.}
In any of two subcases
\seq{\varsigma=2,\ \eta=0\quad \hbox{ or }\quad \varsigma=0,\ \eta=2,} we introduce
\seq{U_k= v_k\quad \hbox{ or }\quad U_k= w_k} and get in any of these four cases the relativistic Toda type equation:
\seq{\ddot U_k=\dot U_k^2\left(\frac{\dot U_{k-1}}{(U_k-U_{k-1})^2}-\frac{\dot U_{k+1}}{(U_k-U_{k+1})^2}+\frac1{U_{k}-U_{k-1}}+\frac1{U_{k}-U_{k+1}}\right),}
where we denote $\dot U_k=\partial_{\tilde\theta_2} U_k$. This is a known equation presented in \cite[Section 3.3.3, (L2) with $r(x,y)=(x-y)^2/2$]{y06}.  

Recently, new examples of five-point differential-difference equations similar to \eqref{sym_nN2} with $a_n\equiv b_n$ and \eqref{s2_tz} have appeared in \cite{a18}.

\subsection{Conservation laws in the $n$-direction}\label{con_law_n}

As equation (\ref{our_bN}) with $N=1$ is well-known, we consider here the cases $N=2,3$ and conservation laws in the $n$-direction, which will be autonomous.

The relation 
\eq{(T_n-1)\check{p}_{n,m}=(T_m-1)\check{q}_{n,m}\label{cln},} with $\check{p}_{n,m},\check{q}_{n,m}$ depending on $n,m,u_{n+i,m+j}$, is called the conservation law of the discrete equation (\ref{our_bN},\ref{bN}) if it is identically satisfied on the solutions of (\ref{our_bN},\ref{bN}). By using equation (\ref{our_bN}), we can rewrite  $\check{p}_{n,m},\check{q}_{n,m}$ in terms of $n,m$ and of the functions \eqref{dym_v} only, and they will be presented in such a form. In case of the $n$-direction, $\check{q}_{n,m}$ will be of the form: $$ \check{q}_{n,m}=\check{q}_{n,m}(u_{n+k_1,m},u_{n+k_1-1,m},\ldots, u_{n+k_2,m}),\quad k_1\geq k_2.$$ The function $\check{q}_{n,m}$ can be called the conserved density by analogy with the discrete-differential case. 

For $k_1> k_2$ we will obtain the densities $\check{q}_{n,m}$, such that $$\dfrac{^2\check{q}_{n,m}}{u_{n+k_1,m}\partial u_{n+k_2,m}}\neq0 \quad\hbox{ for all }\ \  n,m,$$ then the number $k_1-k_2$ will be called the order of such a conservation law. If $k_1=k_2$ and the function $\check{q}_{n,m}$ is not constant, then \eqref{cln} will be the nontrivial zero-order conservation law. Conservation laws of different orders are essentially different.

For the construction of conservation laws in the cases $N=2,3$, we use the master symmetries \eqref{master_nN}. They are simpler than in Section \ref{sec_sym_m} in the sense that do not depend explicitly on the time $\hat\theta_N$ of master symmetry. However, such a way of construction of conservation laws is new even in $\hat\theta_N$-independent case.

We construct a hierarchy of $n$-independent conservation laws for equations (\ref{our_bN},\ref{bN}) with $N=2,3$:
\eq{(T_n-1)\check p_{n,m}^{(k)}=(T_m-1)\check q_{n,m}^{(k)},\quad k\ge 1.\label{pqN}}   Similarly to Section \ref{sec_con_m}, we can construct the conservation laws by induction, using a property that
\eq{(T_n-1)D_{\hat\theta_N} \check p_{n,m}^{(k)} =(T_m-1)D_{\hat\theta_N}\check  q_{n,m}^{(k)}\label{con_tau_n}} is also the conservation law. Here $D_{\hat\theta_N}$ is the total derivative in virtue of \eqref{master_nN}. 

 In both cases $N=2,3$ the starting conservation law will be built with the help of \cite[(33)]{ly09}. Let us denote by \eq{\partial_{\tilde \theta_N}u_{n,m}=\Omega_{n,m}^{(N)}\label{sym_n_auto}} the autonomous symmetries \eqref{sym_nN2} and \eqref{sym_nN3} with $a_n\equiv1$ and rewrite the corresponding discrete equations (\ref{our_bN},\ref{bN}) in the form:
$$u_{n+1,m+1}=\varphi^{(N)}(u_{n,m},u_{n+1,m},u_{n,m+1}).$$ 
Then starting conservation laws read:
\eq{\label{con_n_ly}(1-T_n^N)T_n^{-1}\log \dfrac{\varphi^{(N)}}{u_{n+1,m}}=(T_m-1)\log\dfrac{\Omega_{n,m}^{(N)}}{u_{n+N,m}}.}

Such a conservation law is autonomous with the conserved density $\check q_{n,m}^{(1)}=\log\dfrac{\Omega_{n,m}^{(N)}}{u_{n+N,m}},$ however, the conservation law \eqref{con_tau_n} explicitly depends on the variable $n$. In Section \ref{sec_con_m} it is explained how to remove this dependence on $n$ from such a conservation law. We can do it because the master symmetry \eqref{master_nN} has the linear dependence on $n$, the function $\check q_{n,m}^{(1)}$ is also a conserved density of equation \eqref{sym_n_auto}, and we can add a function of the form \eqref{h12} to both parts of conservation law \eqref{con_tau_n}.

If $N=2$ we can rewrite the conservation law \eqref{con_n_ly} in the form:
\eq{(T_m-1)\check q_{n,m}^{(k)}=(T_n^2-1)\breve p_{n,m}^{(k)},\label{pqN2}} where $k=1$ and
\eq{\check q_{n,m}^{(1)}=\log\frac{(u_{n+1,m}^2-1)(u_{n,m}^2-1)}{U_{n+1,m}^2},\quad\breve p_{n,m}^{(1)}=\log\frac{u_{n,m}+1}{u_{n,m+1}-1},\label{pq22}} while $U_{n,m}$ is defined in \eqref{sym_nN2}.
The form of this conservation law is specific, but it is a particular case of \eqref{pqN} with $\check p_{n,m}^{(k)}=(T_n+1)\breve p_{n,m}^{(k)}.$ By using the master symmetry \eqref{master_nN}, we get the next conservation law which can be made autonomous and rewritten in the same specific form \eqref{pqN2}. It is given by
\seqs{&\check q_{n,m}^{(2)}=-\frac{(u_{n+4,m}+u_{n+3,m})(u_{n+2,m}^2-1)(u_{n+1,m}+u_{n,m})}{U_{n+3,m}U_{n+1,m}}+\frac{u_{n+1,m}^2-1}{U_{n+1,m}},\\&\breve p_{n,m}^{(2)}=\frac{(u_{n+2,m}+u_{n+1,m})(u_{n,m}-1)}{U_{n+1,m}}.}
The orders of these conservation laws are equal to 2 and 4. These conservation laws were constructed in \cite{ghy15} in a little bit different form by using an $L-A$ pair.

If $N=3$  we can rewrite the conservation law \eqref{con_n_ly} in the form:
\eq{(T_m-1)\check q_{n,m}^{(k)}=(T_n^3-1)\breve p_{n,m}^{(k)},\label{pqN3}} where $k=1$, $\breve p_{n,m}^{(1)}$ is defined by \eqref{pq22} as before, and
\seqs{\check q_{n,m}^{(1)}&=\log\frac{(u_{n+2,m}^2-1)(u_{n+1,m}^2-1)(u_{n,m}^2-1)}{U_{n+1,m}^2},} with $U_{n,m}$ defined in \eqref{sym_nN3}. The specific form \eqref{pqN3} is also a particular case of \eqref{pqN}. This conservation law is autonomous and has the order 3. By using the master symmetry \eqref{master_nN}, we can get the next conservation law which can be made autonomous and rewritten in the same specific form \eqref{pqN3} with $k=2$. It will be of the order 6. It is, however, too cumbersome to show it here.

It should be remarked that, using the non-autonomous generalized symmetries \eqref{sym_nN2} and \eqref{sym_nN3} with $a_n\equiv b_n$ and the same formula \eqref{con_n_ly} for starting conservation laws, we can try to get non-autonomous conservation laws. However, nothing new arises because the operator $T_m-1$ annihilates the explicit dependence on $n$.

\subsection{Autonomous $L-A$ pairs}\label{sec_la}
Here we construct autonomous $L-A$ pairs for the discrete equations of series (\ref{our_bN},\ref{bN}), using the non-autonomous $L-A$ pair (\ref{lax_d_dress}-\ref{L2}) for equation \eqref{our_b}. In the case $N=1$, one has $\beta_1=1$ and this $L-A$ pair is obviously autonomous. 

Applying the operator $T_n^{N-1}$ to the first of equations \eqref{lax_d_dress}, we get a consequence:
\eq{\Psi_{n+N,m}=L^{(1,N)}_{n,m}\Psi_{n,m},\quad \Psi_{n,m+1}=L^{(2)}_{n,m}\Psi_{n,m},\label{lax_d_dressN}} where $N\geq2$,  
\eq{L^{(1,N)}_{n,m}=L^{(1)}_{n+N-1,m}L^{(1)}_{n+N-2,m}\ldots L^{(1)}_{n+1,m}L^{(1)}_{n,m},} and $\beta$ is replaced by $\beta_N$ in matrices $L_{n,m}^{(1)},\ L_{n,m}^{(2)}$.  
The compatibility condition for \eqref{lax_d_dressN} is
\eq{L^{(1,N)}_{n,m+1}L^{(2)}_{n,m}=L^{(2)}_{n+N,m}L^{(1,N)}_{n,m}.\label{com_con_lax}}

As $\beta_N^N=1$, we see that the multiplier $\beta_N^n$ is not changed not only in the matrix $L^{(1,N)}_{n,m+1}$ but also in $L^{(2)}_{n+N,m}$.
That is why it plays no role in relation \eqref{com_con_lax}, and we can replace $\beta_N^n$ by a constant. It can be removed by scaling the spectral parameter $\lambda$, and we get the following relation
\eq{\Lambda^{(1,N)}_{n,m+1}\Lambda^{(2)}_{n,m}=\Lambda^{(2)}_{n+N,m}\Lambda^{(1,N)}_{n,m}\label{com_con_lax_auto}}
defined by the matrices: 
\eq{\Lambda^{(1,N)}_{n,m}=\Theta^{(N-1)}_{n,m}\Theta^{(N-2)}_{n,m}\ldots \Theta^{(1)}_{n,m}\Theta^{(0)}_{n,m},}
\eq{\Theta^{(k)}_{n,m}=\matrixx{1&2\lambda\beta_N^k(u_{n+k,m}+1)\\ -\frac2{u_{n+k,m}-1}&\frac{u_{n+k,m}+1}{u_{n+k,m}-1}},}
\eq{\Lambda^{(2)}_{n,m}=\matrixx{1&-\lambda(u_{n,m}+1)(u_{n,m+1}-1)\\ 1&0}.}

For any $N\geq2$ the matrix relation \eqref{com_con_lax_auto} is a consequence of the discrete equation (\ref{our_bN}-\ref{bN}). It turns out that for $N=2,3,4$  relation \eqref{com_con_lax_auto} is equivalent to (\ref{our_bN}-\ref{bN}) as we tested by direct calculation. It is highly likely that the same is true for any $N\geq 2$, and we get  for equation (\ref{our_bN}-\ref{bN}) the following autonomous $L-A$ pair: 
\eq{\Psi_{n+N,m}=\Lambda^{(1,N)}_{n,m}\Psi_{n,m},\quad \Psi_{n,m+1}=\Lambda^{(2)}_{n,m}\Psi_{n,m}\label{lax_d_dressN_auto}.}

\section {Conclusions}

We have constructed a series of autonomous integrable discrete equations (\ref{our_bN}) with $\beta_N$ being a primitive root of unit.  Equation \eqref{our_bN} with $N=1$ is well-known. Equations (\ref{our_bN},\ref{bN}) have hierarchies of autonomous generalized symmetries and conservation laws in both directions as well as autonomous $L-A$ pairs. 

Symmetries and conservation laws of (\ref{our_bN},\ref{bN}) were constructed by using the master symmetries. Those master symmetries arise as generalized symmetries of the discrete equations (\ref{our_bN},\ref{bN}) and linearly depend on one of two discrete variables, see Sections \ref{sec_sym_m},\ref{sym_n}. One of them has also an explicit dependence on its time. In the case of conservation laws, such a construction scheme seems to be new. Entering the time of master symmetry into the corresponding discrete equation also seems to be a new point in the method.

The following hypothesis on the generalized symmetry structure should be true:
\begin{hyp}Any of the autonomous equations (\ref{our_bN},\ref{bN}) has an infinite hierarchy of autonomous generalized symmetries in both directions of the orders $\kappa N,\ \kappa\geq1$. The minimal possible order of an autonomous generalized symmetry in any direction is equal to $N$.
\end{hyp}

We do not know any examples of this kind in case of the hyperbolic partial differential equations which are analogous to discrete equations of the form \eqref{dis_gen}. We justified this hypothesis by results presented in Sections \ref{sym_m_auto},\ref{sym_n}. Corollary~\ref{cor_sym_m} states the existence of autonomous generalized symmetries of orders $\kappa N$ in the $m$-direction. In Theorem \ref{th_sym_n} we have proved, in particular, that equations (\ref{our_bN},\ref{bN}) with $N=2,3$ have autonomous generalized symmetries of the order $N$ in the $n$-direction and do not have autonomous symmetries of lower orders. We can prove a similar result for the $m$-direction:

\begin{theorem}Equations (\ref{our_bN},\ref{bN}) with $N=2,3$ have autonomous generalized symmetries of the order $N$ in the $m$-direction and do not have autonomous symmetries of lower orders. 
\end{theorem}

The proof is similar to one of Theorem \ref{th_sym_n}, but it is too cumbersome to demonstrate here. 

This hypothesis is important from the viewpoint of the generalized symmetry method for discrete equations when we classify discrete equations, using the existence of generalized symmetries of a fixed order \cite{ly11,gy12}. Since the minimal order of autonomous generalized symmetry may be arbitrarily high, we cannot classify in this way all the integrable discrete equations  \eqref{dis_gen} in the autonomous case. 

As our results show, the hierarchies of autonomous conservation laws should have a similar structure. Any of the autonomous equations (\ref{our_bN},\ref{bN}) should have an infinite hierarchy of autonomous conservation laws of the orders $\kappa N,\ \kappa\geq1,$ in both directions. This is true in the $m$-direction, see Corollary \ref{con_m}.

The case $N=2$ is most interesting, as the discrete equation \eqref{our_b2} has no complex coefficients. We consider it in more detail in Section \ref{sec_N2}. For this equation \eqref{our_b2} and its known analogue \eqref{tzit}, we show that their second order generalized symmetries in the $n$-direction have a close relation to integrable differential-difference equations of the relativistic Toda type. We do not know any autonomous discrete  example, except for (\ref{our_b2},\ref{tzit}), with generalized symmetries of this kind.

Finally we remark that, despite the fact that the autonomous equations (\ref{our_bN},\ref{bN}) are more or less obvious particular cases of non-autonomous equations \eqref{our_n}, many results related to these autonomous equations are new. 

\paragraph{Acknowledgments.}  RIY gratefully acknowledges financial support from a Russian Science Foundation grant (project
15-11-20007).

\end{document}